
\input harvmac
\input epsf
\newcount\figno
\figno=0
\def\fig#1#2#3{
\par\begingroup\parindent=0pt\leftskip=1cm\rightskip=1cm\parindent=0pt
\baselineskip=11pt
\global\advance\figno by 1
\midinsert
\epsfxsize=#3
\centerline{\epsfbox{#2}}
\vskip 12pt
{\bf Fig.\ \the\figno: } #1\par
\endinsert\endgroup\par
}
\def\figlabel#1{\xdef#1{\the\figno}}
\def\encadremath#1{\vbox{\hrule\hbox{\vrule\kern8pt\vbox{\kern8pt
\hbox{$\displaystyle #1$}\kern8pt}
\kern8pt\vrule}\hrule}}
\overfullrule=0pt
%
%
\def\tilde{\widetilde}

\def\Z{{\bf Z}}
\def\T{{\bf T}}
\def\S{{\bf S}}
\def\R{{\bf R}}
\def\IA{{\rm IA}}
\def\I{{\rm I}}
\def\IIA{{\rm IIA}}
\def\IIB{{\rm IIB}}

\def\npb#1#2#3{{\it Nucl.\ Phys.} {\bf B#1} (19#2) #3}
\def\plb#1#2#3{{\it Phys.\ Lett.} {\bf B#1} (19#2) #3}

\def\physrev#1#2#3{{\it Phys.\ Rev.} {\bf D#1} (19#2) #3}
\def\ap#1#2#3{{\it Ann.\ Phys.} {\bf #1} (19#2) #3}
\def\physrep#1#2#3{{\it Phys.\ Rep.} {\bf #1} (19#2) #3}

\def\mpla#1#2#3{{\it Mod.\ Phys.\ Lett.} {\bf #1} (19#2) #3}
\def\frac#1#2{{#1\over #2}}

\def\tr{{\rm tr}\,}
\def\semi{\hbox{${}\subset\kern-1em\times{}\,$}}

\font\zfont = cmss10 

\def\bigone{\hbox{1\kern -.23em {\rm l}}}
\def\ZZ{\hbox{\zfont Z\kern-.4emZ}}

\Title{\vbox{\baselineskip12pt
\hbox{hep-th/9510209}
\hbox{IASSNS-HEP-95-86}
\hbox{PUPT-1571}}}
{\vbox{\centerline{HETEROTIC AND TYPE I STRING DYNAMICS}
\bigskip
\centerline{FROM ELEVEN DIMENSIONS}}}
\smallskip
\centerline{ Petr Ho\v rava\footnote{$^\ast$}{horava@puhep1.princeton.edu.
Research supported in part by NSF Grant PHY90-21984.}}
\smallskip
\centerline{\it Joseph Henry Laboratories, Princeton University}
\centerline{\it Jadwin Hall, Princeton, NJ 08544, USA}
\smallskip
\centerline{and}
\smallskip
\centerline{Edward Witten\footnote{$^\star$}{witten@sns.ias.edu.
Research supported in part by NSF Grant PHY92-45317.}}
\smallskip
\centerline{\it School of Natural Sciences, Institute for Advanced Study}
\centerline{\it Olden Lane, Princeton, NJ 08540, USA}\bigskip
\medskip
\noindent
We propose that the ten-dimensional $E_8\times E_8$ heterotic string
is related to an eleven-dimensional theory on the orbifold
${\bf R}^{10}\times {\bf S}^1/{\bf Z}_2$ in the same way that
the Type IIA string in ten dimensions is related to
${\bf R}^{10}\times {\bf S}^1$.  This in particular determines
the strong coupling behavior of the ten-dimensional $E_8\times E_8$
theory.  It also leads to a plausible scenario whereby
duality between $SO(32)$  heterotic and Type I superstrings
follows from the classical symmetries of the eleven-dimensional
world, just as the $SL(2,{\bf Z})$ duality of the ten-dimensional
Type IIB theory follows from eleven-dimensional diffeomorphism invariance.
\Date{October, 1995}
\nref\witten{E. Witten, ``String Theory Dynamics in Various Dimensions,''
\npb{443}{95}{85}, hep-th/9503124.}
\nref\townsend{P.K. Townsend, ``The Eleven-Dimensional Supermembrane
Revisited,'' \plb{350}{95}{184}, hep-th/9501068.}
\nref\duff{M.J. Duff, ``Electric/Magnetic Duality And Its Stringy
Origins,'' Texas A\&M preprint, hep-th/9509106;
M.J. Duff, R.R. Khuri, and J.X. Lu, ``String Solitons,''
\physrep{259}{95}{213}, hep-th/9412184.}
\nref\towndemo{P.K. Townsend, ``$p$-Brane Democracy,'' Cambridge preprint,
hep-th/9507048.}
\nref\schwarz{J.H. Schwarz, ``An $SL(2,Z)$ Multiplet of Type IIB
Superstrings,'' ``Superstring Dualities,'' and ``The Power of M~Theory,''
Caltech preprints, hep-th/9508143, 9509148, and 9510086.}
\nref\supermemb{E. Bergshoeff, E. Sezgin and P.K. Townsend, ``Supermembranes
and Eleven Dimensional Supergravity,'' \plb{189}{87}{75}; ``Properties of the
Eleven-Dimensional Supermembrane Theory,'' \ap{185}{88}{330}.}
\nref\dab{A. Dabholkar, ``Ten-Dimensional Heterotic String As A Soliton,''
\plb{357}{95}{307}, hep-th/9506160.}
\nref\hull{C. M. Hull, ``String-String Duality In Ten Dimensions,''
\plb{357}{95}{545}, hep-th/9506194.}
\nref\pw{J. Polchinski and E. Witten, ``Evidence For Heterotic - Type I
String Duality,'' IAS/ITP preprint, hep-th/9510169.}
\nref\aspinwall{P.S. Aspinwall, ``Some Relationships Between Dualities in
String Theory,'' Cornell preprint, hep-th/9508154.}
\nref\alv{L. Alvarez-Gaum\'e and E. Witten, ``Gravitational Anomalies,''
\npb{234}{83}{269}.}
\nref\milnor{M. Kervaire and J. Milnor, ``Groups of Homotopy Spheres I,''
{\it Ann.\ Math.} {\bf 77} (1963) 504.}
\nref\gs{M.B. Green and J.H. Schwarz, ``Anomaly Cancellations in
Supersymmetric $D=10$ Gauge Theory and Superstring Theory,''
\plb{149}{84}{117}.}
\nref\double{M.J. Duff, P.S. Howe, T. Inami and K.S. Stelle, ``Superstrings
in $D=10$ from Supermembranes in $D=11$,'' \plb{191}{87}{70}.}
\nref\hughes{J. Hughes and J. Polchinski, ``Partially Broken Global
Supersymmetry and the Superstring,'' \npb{278}{86}{147}.}
\nref\duffo{M.J. Duff, J.T. Liu and R. Minasian, ``Eleven Dimensional
Origin of String/String Duality: A One Loop Test,'' Texas A\&M preprint,
hep-th/95061126.}
\nref\blum{J. Blum and J.A. Harvey, ``Anomaly Inflow for Gauge Defects,''
\npb{416}{94}{119}.}
\nref\bergsh{E. Bergshoeff, C. Hull and T. Ort\'\i n, ``Duality in the
Type-II Superstring Effective Action,'' \npb{451}{95}{547}, hep-th/9504081;
E.~Bergshoeff, B.~Janssen and T.~Ort\'\i n, ``Solution-Generating
Transformations and the String Effective Action,'' hep-th/9506156.}
\nref\sagnotti{A. Sagnotti, ``Open Strings and Their Symmetry Groups,'' in:
Carg\`ese '87, ``Nonpereturbative Quantum Field Theory,'' eds.: G. Mack et
al.\ (Pergamon Press,1988) p.\ 521}
\nref\wso{P. Ho\v rava, ``Strings on World-Sheet Orbifolds,''
\npb{327}{89}{461}.}
\nref\polch{J. Dai, R.G. Leigh and J. Polchinski, ``New Connections Between
String Theories,'' \mpla{4}{89}{2073}.}
\nref\wsodual{P. Ho\v rava, ``Background Duality of Open String Models,''
\plb{231}{89}{351}.}
\nref\poldir{J. Polchinski, ``Dirichlet-Branes and Ramond-Ramond Charges,''
ITP preprint, hep-th/9510017.}
\nref\cso{P. Ho\v rava, ``Chern-Simons Gauge Theory on Orbifolds: Open
Strings from Three Dimensions,'' hep-th/9404101.}
\nref\otherpolch{J. Polchinski, ``Combinatorics of Boundaries In String
Theory,''\physrev{50}{94}{6041}, hep-th/9407031.}
\nref\narain{K.S. Narain, ``New Heterotic String Theories in Uncompactified
Dimensions $<10$,'' \plb{169}{86}{41}; K.S. Narain, M.H. Sarmadi and E.
Witten, ``A Note on Toroidal Compactification of Heterotic String Theory,''
\npb{279}{87}{369}.}
\nref\ginsparg{P. Ginsparg, ``On Toroidal Compactification of Heterotic
Superstrings,'' \physrev{35}{87}{648}.}
\newsec{Introduction}

In the last year, the strong coupling behavior of many supersymmetric string
theories (or more exactly of what we now understand to be the one
supersymmetric string theory in many of its simplest vacua) has been
determined.  For instance, the strong coupling behavior of most of the
ten-dimensional theories and their toroidal compactifications seems to be
under control \witten.
A notable exception is the $E_8\times E_8$ heterotic
string theory in ten dimensions; no proposal has yet been made that would
determine its low energy excitations and interactions in the strong coupling
regime.  One purpose of this paper is to fill this gap.

We also  wish to further explore the relation of string theory to eleven
dimensions.  The strong coupling behavior of the Type IIA theory in ten
dimensions has turned out \refs{\townsend,\witten} to involve
eleven-dimensional supergravity on $\R^{10}\times\S^1$, where the radius of
the $\S^1$ grows with the string coupling.  An eleven-dimensional
interpretation of string theory has had other applications, some of them
explained in \refs{\duff - \schwarz}.  The most ambitious interpretation of
these facts is to suppose that there really is a yet-unknown
eleven-dimensional quantum theory that underlies many aspects of string
theory, and we will formulate this paper as an exploration of that theory.
(But our arguments, like some of the others that have been given, could be
compatible with interpreting the eleven-dimensional world as a limiting
description of the low energy excitations for strong coupling, a view taken
in \witten.)  As it has been proposed that the eleven-dimensional theory is
a supermembrane theory but there are some reasons to doubt that
interpretation,%
\foot{To get the right spectrum of BPS saturated states after toroidal
compactification, the eleven-dimensional theory should support stable
macroscopic membranes of some sort, presumably described at long wavelengths
by the supermembrane action \refs{\supermemb,\townsend}.  We will indeed make
this assumption later.  But  that the theory can be understood as a theory of
fundamental membranes seems doubtful because (i) on the face of it, membranes
cannot be quantized; (ii) there is no dilaton or coupling parameter that
would justify a classical expansion in membranes.}
we will non-committally call it the $M$-theory, leaving to the future
the relation of $M$ to membranes.

Our approach to learning more about the $M$-theory is to consider its
behavior on a certain eleven-dimensional orbifold ${\bf R}^{10}\times
{\bf S}^1/{\bf Z}_2$.  In the process, beyond  making a proposal for how
the $E_8\times E_8$ heterotic string is related to the $M$-theory, we will
make a proposal for relating the classical symmetries of the $M$-theory to
the conjectured heterotic - Type I string duality in ten dimensions
\refs{\witten,\dab -\pw}, much as the classical symmetries of the $M$-theory
have been related to Type II duality symmetries \refs{\witten,\schwarz,
\aspinwall}.  These proposals suggest a common eleven-dimensional origin of
all ten-dimensional string theories and their dualities.

\newsec{The $M$-Theory On An Orbifold}

The $M$-theory has for its low energy limit eleven-dimensional
supergravity.  On an eleven-manifold,  with signature $-++\dots +$, we
introduce gamma matrices $\Gamma^I$, $I=1,\dots, 11$, obeying
$\{\Gamma_I,\Gamma_J\}=2\eta_{IJ}$ and (in an oriented orthonormal
basis)
\eqn\gur{\Gamma^1\Gamma^2\cdots
\Gamma^{11}=1.}
We will assume that the $M$-theory has enough in
common with what we know of string theory that it makes sense
on a wide class of orbifolds -- but possibly, like string theory,
with extra massless modes arising at fixed points.  We will
consider the $M$-theory on the particular orbifold ${\bf R}^{10}\times
{\bf S}^1/{\bf Z}^2$, where ${\bf Z}_2$ acts on ${\bf S}^1$ by
$x^{11}\to -x^{11}$, reversing the orientation.  Note that
eleven-dimensional supergravity is invariant under
orientation-reversal if accompanied by change in sign of the
three-form $A^{(3)}$, so this makes sense at least for the massless
modes coming from eleven dimensions.%
\foot{One might wonder whether there is a global anomaly that spoils parity
conservation, as described on p.\ 309 of \alv.  This does not occur, as
there are no exotic twelve-spheres \milnor.}

On $\R^{10}\times \S^1$, the $M$-theory is invariant under supersymmetry
generated by an arbitrary constant spinor  $\epsilon$.
Dividing by ${\bf Z}_2$ kills half the supersymmetry; sign conventions
can be chosen so that the unbroken supersymmetries are     generated
by constant spinors $\epsilon$ with $\Gamma^{11}\epsilon=\epsilon$.
Together with \gur, this condition means that
\eqn\nur{\Gamma^1\Gamma^2\cdots\Gamma^{10}\epsilon=\epsilon,}
so $\epsilon$ is chiral in the ten-dimensional sense.

The $M$-theory on $\R^{10}\times \S^1/\Z_2$ thus reduces at low
energies to a ten-dimensional Poincar\'e-invariant
supergravity theory with one chiral supersymmetry.
There are three string theories with that low energy structure,
namely the  $E_8\times E_8$ heterotic string and the two
theories -- Type I and heterotic -- with $SO(32)$ gauge group.
It is natural to wonder whether the $M$-theory on $\R^{10}\times
\S^1/\Z_2$ reduces, as the radius of the $\S^1$ shrinks to zero,
to one of these three string theories, just as the $M$-theory
on $\R^{10}\times \S^1$ reduces to the Type IIA superstring in the
same limit.  We will give three arguments that all show that if the
$M$-theory on $\R^{10}\times \S^1/\Z_2$ reduces for small radius to
one of the three string theories, it must be the $E_8\times E_8$
heterotic string.  The arguments are based respectively
on space-time gravitational anomalies, the strong coupling behavior,
and world-volume gravitational anomalies.

\bigskip\noindent
{\it (i) Gravitational Anomalies}
\par\nobreak\smallskip\nobreak

First we consider the gravitational anomalies of the $M$-theory
on $\R^{10}\times \S^1/\Z_2$; these
 should be computable  without detailed knowledge of the $M$-theory
because anomalies can be computed from only a knowledge of the
low-energy structure.
In raising the question, we understand
a metric on $\R^{10}\times \S^1/\Z_2$ to be a metric on $\R^{10}\times
\S^1$ that is invariant under $\Z_2$; a diffeomorphism of
$\R^{10}\times \S^1/\Z_2$ is a diffeomorphism of $\R^{10}\times
\S^1$ that commutes with $\Z_2$.  The standard massless fermions
of the $M$-theory are the gravitinos; by a gravitino mode on
$\R^{10}\times\S^1/\Z_2$ we mean a gravitino mode on $\R^{10}\times \S^1$
that is invariant under $\Z_2$.  With these specifications,
it makes sense to ask whether the effective action obtained by
integrating out the gravitinos on $\R^{10}\times \S^1/\Z_2$
is anomaly-free, that is, whether it is invariant under diffeomorphisms.

First of all, on a smooth
eleven-manifold, the effective action obtained by integrating
out gravitinos is anomaly-free; purely gravitational anomalies
are absent (except possibly for
global anomalies in $8k$ or $8k+1$ dimensions)
 in any dimension not of the form $4k+2$ for some integer
$k$.  But the result on an orbifold is completely different.
In the case we are considering, it is immediately apparent that
the Rarita-Schwinger field has a gravitational anomaly.  In fact,
the eleven-dimensional Rarita-Schwinger field reduces in ten
dimensions to a sum of infinitely many massive fields (anomaly-free)
and the massless chiral ten-dimensional gravitino discussed
above -- which \alv\ gives an anomaly under
ten-dimensional diffeomorphisms. Thus, at least under those
diffeomorphisms of the eleven-dimensional orbifold that come
from diffeomorphisms of $\R^{10}$ (times the trivial diffeomorphism
of $\S^1/\Z_2$), there is an anomaly.

To compute the  form of this anomaly, it is not necessary to do
anything essentially new; it is enough to know the standard
Rarita-Schwinger anomaly on a ten-manifold, as well as the result
(zero) on a smooth eleven-manifold.  Thus, under a space-time
diffeomorphism $\delta x^I=\epsilon v^I$ generated by a vector
field $v^I$, the change of the effective action is on general
grounds of the form
\eqn\ochange{\delta\Gamma =i\epsilon
\int_{\R^{10}\times \S^1/\Z_2}d^{11}x\sqrt g
 v^I(x) W_I(x),}
where $g$ is the eleven-dimensional metric
and $W_I(x) $ can be computed {\it locally}
from the data at $x$.  The existence of a local expression for $W_I$
reflects the fact that the anomaly can be understood to result
entirely from failure of the regulator to preserve the symmetries,
and so can be computed from short distances.

Now, consider the possible form of $W_I(x)$ in our problem.
If $x$ is a {\it smooth} point in $\R^{10}\times \S^1/\Z_2$
(not an orbifold fixed point), then lack of anomalies of the
eleven-dimensional theory implies that $W_I(x)=0$.  $W_I$ is
therefore a sum of delta functions supported on the fixed
hyperplanes $x^{11}=0$ and $x^{11}=\pi$, which we will call $H'$ and
$H''$,  so \ochange\ actually
takes the form
\eqn\jchange{\delta\Gamma=i\epsilon\int_{H'}d^{10}x\sqrt{g'}
v^IW_I'+i\epsilon \int_{H''}d^{10}x\sqrt{g''}
v^IW_I''}
where now $g'$ and $g''$ are the restrictions of $g$ to $H'$
and $H''$  and $W'$, $W''$ are local functionals constructed
from the data on those hyperplanes.  Obviously, by symmetry,
$W''$ is the same as $W'$, but defined from the metric at
$H''$ instead of $H'$.
The form of $W'$ and $W''$ can be determined as follows without
any computation.  Let the metric on $\R^{10}\times \S^1/\Z^2$ be
the product of an arbitrary metric on $\R^{10}$ and a standard
metric on $\S^1/\Z_2$, and take $v$ to be the pullback of a vector
field on $\R^{10}$.  In this situation (as we are simply studying
a massless chiral gravitino on $\R^{10}$ plus infinitely many
massive fields), $\delta\Gamma$ must simply equal the standard
ten-dimensional anomaly.  The two contributions in \jchange\ from
$x^{11}=0$ and $x^{11}=\pi$ must therefore each give {\it one-half}
of the standard ten-dimensional answer.  Though we considered
a rather special configuration to arrive at this result, it was
general enough to permit an arbitrary metric at $x^{11}=0$
(or $\pi$) and hence to determine the functionals $W'$, $W''$ completely.

Since the anomaly is not zero, the massless modes we know about
cannot be the whole story for the $M$-theory on this orbifold.
There will have to be additional massless modes that propagate
only on the fixed planes; they will be analogous to the twisted
sector modes of string theory orbifolds.  The modes will have to
be ten-dimensional vector multiplets because the vector multiplet
is the only ten-dimensional supermultiplet with all spins $\leq 1$.
Let us determine what vector multiplets there may be.

First of all, part of the anomaly can be canceled by a generalized
Green-Schwarz mechanism \gs , with the fields $B'$ and $B''$, defined as
the components $A^{(3)}_{ij 11}$ of the three-form on $H'$ and $H''$,
entering roughly as the usual $B$ field does in the
Green-Schwarz mechanism.  There will be interactions
$\int_{H'} B'\wedge Z'_8$
and $\int_{H''}B''\wedge Z''_8$
at the fixed planes, with some eight-forms $Z'$ and $Z''$,
and in addition the gauge transformation
law of $A^{(3)}$  will have terms proportional to delta functions
supported on $H'$ and $H''$.  In this way -- as in the more
familiar ten-dimensional case -- some of the anomalies can be
canceled, but not all.

In fact, recall that the anomaly in ten-dimensional supergravity
is constructed from a twelve-form $Y_{12}$ that
is a linear combination of  $(\tr R^2)^3$, $\tr R^2\cdot \tr R^4$,
and $\tr R^6$ (with $R$ the curvature two-form and  $\tr$ the trace
in the vector representation).  The first two terms
are ``factorizable'' and can potentially be canceled by a Green-Schwarz
mechanism.  The last term is ``irreducible'' and cannot be so
canceled.  The irreducible part of the anomaly must be canceled
by additional massless modes -- necessarily vector multiplets -- from
the ``twisted sectors.''

In ten dimensions, the story is familiar \gs.
The irreducible part of the standard ten-dimensional anomaly can
 be canceled precisely by the addition of     496 vector
multiplets, so that the possible gauge groups in ten-dimensional
$N=1$ supersymmetric string theory have dimension 496.  We are in the
same situation now except that the standard anomaly is divided
equally between the two fixed hyperplanes.  We must have therefore
precisely 248 vector multiplets propagating on each of the
two hyperplanes! 248 is, of course, the dimension of $E_8$.
So if the $M$-theory on this orbifold is to
be related to one of the three string theories, it must be the
$E_8\times E_8$ theory, with one $E_8$ propagating on each hyperplane.
$SO(32) $ is not possible as gauge invariance would force us to put
all the vector multiplets on one hyperplane or the other.

By placing one $E_8$ at each end, we cancel the irreducible part of
the anomaly, but it may not be immediately apparent that the reducible
part of the anomaly can be similarly canceled.  To see that this
is so, recall some facts about the standard ten-dimensional anomaly.
With the gauge fields  included, the anomaly is derived from
a twelve-form $\tilde Y_{12}$ that is a polynomial in $\tr F_1^2$ and
$\tr F_2^2$ ($F_1$ and $F_2$ are
the two $E_8$ curvatures; the symbol $\tr $ denotes $1/30$ of the
trace in the adjoint representation) as well as $\tr R^2$,
$\tr R^4$.  ($\tr R^6$ is absent as that part has been canceled by
adding vector multiplets.)
  $\tilde Y_{12}$ has the properties
\eqn\undp{\eqalign{{\partial^2\tilde Y_{12}\over \partial\,\tr F_1^2
{}~\partial\,\tr F_2^2}
 & = 0 \cr
\tilde Y_{12}  = \left(\tr F_1^2+\right.&\left.
\tr F_2^2-\tr R^2\right) \wedge \tilde
Y_8\cr}}
where the details of the polynomial $\tilde Y_8$ will not be essential.
The factorization in the second  equation is the key to anomaly
cancellation.  The first equation reflects the fact that (as
the massless fermions are in the adjoint representation) there is
no massless fermion charged under each $E_8$, so that the anomaly
has no ``cross-terms'' involving both $E_8$'s.

Note that if we set $U_i=\tr F_i^2-{1\over 2}\tr R^2$ for $i=1,2$,
then the first equation in \undp\ implies that
\eqn\dudu{\eqalign{
U_1\wedge &\left( \tilde Y_8(U_1,U_2,\tr R^2,\tr R^4) -\tilde Y_8(U_1,0,
\tr R^2,\tr R^4\right)\cr +U_2\wedge
& \left(\tilde Y_8(U_1,U_2,\tr R^2,\tr R^4) -\tilde Y_8(0,U_2,
\tr R^2,\tr R^4)\right) =0.\cr}}  Hence we can write
\eqn\jundp{\eqalign{\tilde Y_{12} =& (\tr F_1^2 -{1\over 2}\tr R^2) \wedge
 Z_8(\tr F_1^2,\tr R^2,\tr R^4) \cr&+
 (\tr F_2^2 -{1\over 2}\tr R^2) \wedge
 Z_8(\tr F_2^2,\tr R^2,\tr R^4).\cr}}
Here $Z_8$ is defined by $Z_8 (\tr F_1^2,\tr R^2,\tr R^4)=
\tilde Y_8(U_1,0,\tr R^2,\tr R^4)$.  \jundp\ is the desired
formula showing how the anomalies can be canceled by a variant
of the Green-Schwarz mechanism adapted to the eleven-dimensional
problem.  The first term, involving $F_1$ but not $F_2$, is the
contribution from couplings of what was above called
$B'$, and the second term, involving $F_2$ but not $F_1$, is the
contribution from $B''$.

\bigskip\noindent
{\it (ii) Strong Coupling Behavior}
\par\nobreak\smallskip\nobreak

If it is true that the $M$-theory on this orbifold is related
to the $E_8\times E_8$ superstring theory, then the relation
between the radius $R$ of the circle (in the eleven-dimensional
metric) and the string coupling constant $\lambda$ can be determined
by comparing the predictions of the two theories for
 the low energy effective action of the supergravity
multiplet in ten dimensions.  The analysis is precisely as in
\witten\ and will not be repeated here.  It gives the same
relation
\eqn\reldet{ R=\lambda^{2/3}}
that one finds between the $M$-theory
on $\R^{10}\times \S^1$ and Type IIA superstring theory.

In particular, then, for small $R$ -- where the supergravity
cannot be a good description as $R$ is small compared to the Planck
length -- the string theory is weakly coupled and can be a good
description.  On the other hand, we get a candidate for the
strong coupling behavior of the $E_8\times E_8 $ heterotic string:
it corresponds to supergravity on    the $\R^{10}\times {\bf S}^1/
{\bf Z}_2$ orbifold, which is an effective description
(of the low energy interactions of the light modes) for large
$\lambda$  as then $R$ is much bigger than the Planck length.
If our proposal is correct, then
what a low energy observer sees in the strongly coupled $E_8\times E_8$
theory depends on where he or she is; a generic observer, far from
one of the fixed hyperplanes, sees simply eleven-dimensional supergravity
(or the $M$-theory), and does not distinguish the strongly coupled
$E_8\times E_8$ theory from a strongly coupled Type IIA theory,
 while an observer near one of the distinguished hyperplanes
sees eleven-dimensional supergravity on a half-space, with
an $E_8$ gauge multiplet propagating on the boundary.

We can also now see another reason that if the $M$-theory on
${\bf R}^{10}\times {\bf S}^1/{\bf Z}_2$ is related to one of the
three ten-dimensional string theories, it must be the $E_8\times
E_8$ heterotic string.  Indeed, there is by now convincing evidence
\refs{\witten, \dab - \pw} that the strong coupling limit of the
Type I superstring in ten dimensions is the weakly coupled
$SO(32) $ heterotic string, and vice-versa, so we would not want
to relate either of the two $SO(32)$ theories to eleven-dimensional
supergravity.  We must relate the orbifold to the $E_8\times E_8$
theory whose strong coupling behavior has been previously unknown.

\bigskip\noindent
{\it (iii) Extended Membranes}
\par\nobreak\smallskip\nobreak

As our third and last piece of evidence, we want to consider extended
membrane states in the $M$-theory after further compactification
to $\R^9\times \S^1\times \S^1/\Z_2$.

Our point of view is not that the $M$-theory ``is'' a theory of
membranes but that it describes, among other things, membrane states.
There is a crucial difference.  For instance, any spontaneously
broken unified gauge theory in four dimensions with an unbroken
$U(1)$ describes, among other things, magnetic monopoles.  That
does not mean that the theory can be recovered by quantizing
magnetic monopoles; that is presumably possible
only in very special cases.  Classical magnetic monopole solutions
of gauge theory, because of their topological stability, can be
quantized to give quantum states.  But topologically
unstable monopole-antimonopole configurations,
while representing possibly interesting
classical solutions, cannot ordinarily be quantized to understand
photons and electrons.  Likewise, we assume that when the topology
is right, the $M$-theory has topologically stable membranes
(presumably described if the length scale is large enough
by the low energy supermembrane action \townsend) that
can be quantized to give quantum eigenstates.  Even when the topology is
wrong -- for instance on $\R^{11}$ where there is no    two-cycle for
the membrane to wrap around -- macroscopic membrane solutions
(with a scale much bigger than the Planck scale) will make sense,
but we do not assume that they can be quantized to recover gravitons.

The most familiar example of a situation in which there are topologically
stable membrane states is that of compactification of the
$M$-theory on $\R^9\times \S^1\times \S^1$.  With $x^1$ understood
as the time and $x^{10},x^{11}$ as the two periodic variables,
the classical membrane equations have a solution described
by $x^2=\dots = x^9=0$.  This solution is certainly topologically stable
so (if the radii of the circles are big in Planck units) it can be
reliably quantized to obtain quantum states.  The solution
is invariant under half of the supersymmetries,
namely those obeying
\eqn\utty{\Gamma^1\Gamma^{10}\Gamma^{11}\epsilon = \epsilon,}
so these will
be BPS-saturated states. This latter fact gives the quantization
of this particular membrane solution a robustness that enables one
(even if the membrane in question can {\it not} for other purposes
be usefully treated as elementary)
to extrapolate to a regime in which one of the $\S^1$'s is small
and one can compare to weakly coupled string theory.

Let us recall the result
of this comparison, which goes under the name of double dimensional
reduction of the supermembrane \double .
The membrane solution described above breaks the eleven-dimensional
Lorentz group to $SO(1,2) \times SO(8)$.  The massless modes
on the membrane world-volume are the  oscillations of
$x^2,\dots, x^9$, which transform as $({\bf 1},{\bf 8})$,
and fermions that transform as $({\bf 2},{\bf 8}'')$.  Here  ${\bf 2}$ is
the spinor of $SO(1,2)$ and ${\bf 8}$, ${\bf 8}'$, and ${\bf 8}''$
are the vector and the two spinors of $SO(8)$.  To interpret this
in string
theory terms, one considers only the zero modes in the $x^{11}$ direction,
and decomposes the spinor of $SO(1,2)$ into positive and negative chirality
modes of $SO(1,1)$; one thus obtains  the world-sheet structure of the
Type IIA superstring.  So those  membrane excitations that
are low-lying when the second circle is small will match up properly
with Type IIA  states.  Since many of these Type IIA states
are BPS saturated and can be followed from   weak to strong coupling,
the membrane we started with was really needed to reproduce this part of the
spectrum.

Now we move on the the $\R^{10}\times \S^1/\Z_2$ orbifold.
We assume that the classical membrane solution $x^2=\dots = x^9$ is
still allowed on the orbifold; this amounts to assuming that the
membranes of the $M$-theory can have boundaries that lie on the
fixed hyperplanes.  In the orbifold, unbroken supersymmetries
(as we discussed at the outset of section two) correspond to spinors
$\epsilon$ with $\Gamma^{11}\epsilon=\epsilon$;
these transform as the ${\bf 16}$ of $SO(1,9)$, or as ${\bf 8}'_+\oplus
{\bf 8}''_-$ of $SO(1,1)\times SO(8)$.
 The spinors unbroken
in the field of the membrane solution also obey \utty, or equivalently
$\Gamma^1\Gamma^{10}\epsilon=\epsilon$.  Thus, looking at the situation
in string terms (for an observer who does not know about the eleventh
dimension), the unbroken supersymmetries have positive chirality on the
string world sheet and transform as ${\bf 8}'_+$
(where the $+$ is the $SO(1,1)$ chirality) under $SO(1,1)\times SO(8)$.
The massless world-sheet bosons, oscillations in $x^2,\dots,x^9$, survive
the orbifolding, but half of the fermions are projected out.
The survivors transform as ${\bf 8}''_-$; one can think of them \hughes\
 as Goldstone fermions for the
${\bf 8}''_-$ supersymmetries that are broken by the classical membrane
solution.  The $-$ chirality means that they are right-moving.

So the massless modes we know about transform like the world-sheet modes
of the heterotic string that carry space-time quantum numbers:
left and right-moving bosons transforming in the ${\bf 8}$ and right-moving
fermions in the ${\bf 8}''$.  Recovering much of the world-sheet structure
of the heterotic string does not imply that the string theory (if any)
 related
to the $M$-theory
orbifold is a heterotic rather than Type I string; the Type I
theory also describes among other things an object with the world-sheet
structure of the heterotic string \pw.  It is by considerations of
anomalies on the membrane world-volume that we will reach an interesting
conclusion.

The Dirac operator on the membrane three-volume is free of world-volume
gravitational anomalies as long as the world-volume is a smooth manifold.
(Recall that except possibly for discrete anomalies in $8k$ or $8k+1$
dimensions, gravitational anomalies occur only in dimensions $4k+2$.)
In the present case, the world-volume is not a smooth manifold, but
has orbifold singularities (possibly better thought of as boundaries)
at $x^{11}=0$ and at $x^{11}=\pi$.  These singularities give rise
to  three-dimensional
gravitational anomalies; this is obvious from the fact that the massless
world-sheet fermions in the two-dimensional sense
are the fermions ${\bf 8}''_-$ of definite chirality.
By analysis just like we gave in the eleven-dimensional case, the
gravitational anomaly on the membrane world-volume
 vanishes at smooth points
and is a sum of delta functions supported at $x^{11}=0$ and $x^{11}=\pi$.
As in the eleven-dimensional case, these delta functions each represent
{\it one half} of the usual two-dimensional anomaly of the
effective massless two-dimensional ${\bf 8}''_-$ field.

As is perhaps obvious intuitively and we will argue below, the gravitational
anomaly of the ${\bf 8}''_-$ field is the usual gravitational anomaly
of right-moving RNS fermions and superconformal ghosts in the heterotic
string.  So far we have only considered the modes that propagate in bulk
on the membrane world-volume.  If the membrane theory makes sense
in the situation we are considering, the gravitational anomaly
of the ${\bf 8}''_-$ field must be canceled by   additional
world-volume ``twisted sector'' modes, supported at the
 orbifold fixed points $x^{11}=0$ and
$x^{11}=\pi$.  If we are to recover one of the
known string theories, these ``twisted sector modes''
should be left-moving current algebra modes
with $c=16$.  (In any event there is practically no other way to maintain
space-time supersymmetry.)  Usually both $SO(32) $ and $E_8\times E_8$
are possible, but in the present context the anomaly that must be cancelled
is supported one half at $x^{11}=0 $ and one half at $x^{11}=\pi$,
so the only possibility is to have $E_8\times E_8$ with one $E_8$
supported at each end.  This then is our third reason that if the
$M$-theory on the given orbifold has a known string theory as its weak
coupling limit, it must be the $E_8\times E_8$ heterotic string.

It remains to discuss somewhat more carefully the gravitational
anomalies that have just been exploited.  Some care is needed here
as there is an  important distinction between objects that are quantized
as elementary strings and objects that are only known as macroscopic
strings embedded in space-time.  (See, for instance, \refs{\duffo,\blum}.)
For elementary strings, one usually
considers separately both right-moving and left-moving conformal anomalies.
The sum of the two is the total conformal anomaly (which generalizes to
the conformal anomaly in dimensions above two), while the difference
is the world-sheet gravitational anomaly.  For objects that are only
known as macroscopic strings embedded in space-time, the total conformal
anomaly is
not a natural concept, since the string world-sheet has a natural
metric (and not just a conformal class of metrics) coming from the embedding.
But the gravitational anomaly, which was exploited above, still makes sense
even in this situation, as the world-sheet is still not endowed with a
natural coordinate system.

Let us justify the claim that the world-sheet gravitational anomaly of the
${\bf 8}''_-$ fermions encountered above equals the usual gravitational
anomaly from right-moving RNS fermions and
superconformal ghosts.  A detailed calculation is not necessary, as this
can be established by the following simple means.  First let
us state the problem
(as it appears after double dimensional reduction to the Green-Schwarz
formulation of the heterotic string) in generality.
The problem
 really involves, in general, a two-dimensional world-sheet $\Sigma$ embedded
in a ten-manifold $M$.  The normal bundle $N$ to the world-sheet is
a vector bundle with structure group $SO(8)$.  If $S_-$ is the bundle
of negative chirality spinors on $\Sigma$ and $N''$ is the bundle associated
to $N$ in the ${\bf 8}''$ representation of $SO(8)$, then the ${\bf 8}''_-$
fermions that we want are sections of $L_-\otimes N''$.  By making
a triality transformation in $SO(8)$,
 we can replace the fermions with sections of $L_-\otimes N$ without
changing the anomalies.  Now using the fact that the tangent bundle
of $M$ is the sum of $N$ and the tangent bundle of $\Sigma$ -- $TM=N\oplus
T\Sigma$ -- we can replace $L_-\otimes N$ by $L_-\otimes TM$ if we also
subtract the contribution of fermions that take values in
$L_-\otimes T\Sigma$.  The $L_-\otimes TM$-valued fermions are the usual
right-moving RNS fermions, and (as superconformal ghosts take values in
$L_-\otimes T\Sigma$) subtracting the contribution of fermions valued in
$L_-\otimes T\Sigma$ has the same effect as including the superconformal
ghosts.

\newsec{Heterotic - Type I Duality from the $M$-Theory}

In this section, we will try to relate the eleven-dimensional
picture to another interesting phenomenon, which is the conjectured
duality between the heterotic and Type I $SO(32)$ superstrings.

So far we have presented arguments indicating that the $E_8\times E_8$
heterotic string theory is related to the $M$-theory on
$\R^{10}\times\S^1/\Z_2$, just as the Type IIA theory is related to the
$M$-theory compactifed on $\R^{10}\times\S^1$.

We can follow this analogy one step further, and
compactify the tenth dimension of the $M$-theory on $\S^1$.
Schwarz \schwarz\ and Aspinwall \aspinwall\ explained how the $SL(2,\Z)$
duality of the ten-dimensional Type IIB string theory follows from space-time
diffeomorphism symmetry of the $M$-theory on $\R^9\times\T^2$.  (For some
earlier results in that direction see also \bergsh.)  Here we will argue that
the $SO(32)$ heterotic - Type I duality similarly follows from classical
symmetries of the $M$-theory on $\R^9\times\S^1\times\S^1/\Z_2$.

First we need several facts about $T$-duality of open-string models.

\bigskip\noindent
{\it T-Duality in Type I Superstring Theory}
\par\nobreak\smallskip\nobreak

The Type I theory in ten dimensions can be interpreted as a generalized
$\Z_2$ orbifold of the Type IIB theory \refs{\sagnotti - \poldir}.  The
orbifold in question acts by reversing world-sheet parity, and acts trivially
on the space-time.  Projection of the Type IIB spectrum to $\Z_2$-invariant
states makes the Type IIB strings unoriented; this creates an anomaly in the
path integral over world-sheets with crosscaps, which must be compensated for
by introducing boundaries.  The open strings, which are usually introduced to
cancel the anomaly, are naturally interpreted as the twisted states of the
parameter-space orbifold.

More generally, one can combine the reversal of world-sheet orientation
with a space-time symmetry, getting a variant of the Type I theory \refs{\wso
- \wsodual}.%
\foot{More generally still, one can divide by a group containing
some elements that act only on space-time and some that also reverse
the world-sheet orientation; the construction of the twisted states
then has certain subtleties that were discussed in \cso.}
A special case of this will be important here.
Upon compactification to $\R^9\times \S^1$, Type IIB theory is $T$-dual to
Type IIA theory.  Analogously, the Type I theory -- which is a $\Z_2$
orbifold of Type IIB theory -- is $T$-dual to a certain $\Z_2$ orbifold of
Type IIA theory.
This orbifold is constructed by dividing the Type IIA theory
 by a $\Z_2$ that reverses
 the world-sheet orientation  and acts on the circle by
$x^{10}\to -x^{10}$.
Note that the Type IIA theory is invariant under combined reversal
of world-sheet and space-time orientations (but not under either
one separately), so the combined operation is a symmetry.
This theory has been called the Type I$'$ or Type IA theory.
In this theory, the twisted states are open strings that have
their endpoints at the fixed points $x^{10}=0$ and $x^{10}=\pi$.
To cancel anomalies, these open strings must carry Chan-Paton factors.
If we want to treat the two fixed points symmetrically -- as in natural in
an orbifold -- while canceling the anomalies, there must be  $SO(16)$
Chan-Paton factors at each fixed point, so the gauge group is
$SO(16)\times SO(16)$.

In fact, it has been shown \refs{\wso - \wsodual} that the $SO(16)\times
SO(16)$ theory just described is the $T$-dual of the vacuum of the
standard Type I theory on $\R^9\times \S^1$ in which $SO(32)$ is
broken to $SO(16)\times SO(16)$ by a Wilson line.  This is roughly
because $T$-duality exchanges the usual Neumann boundary conditions
of open strings with Dirichlet boundary conditions, and gives a theory
in which the open strings terminate at the fixed points.

Of course, the Type I theory on $\R^9\times \S^1$ has moduli corresponding
to Wilson lines; by adjusting them one can change the unbroken gauge
group or restore the full $SO(32)$.  In the $T$-dual description,
turning on these moduli causes the positions at which the open strings
terminate to vary in a way that depends upon their Chan-Paton charges
\otherpolch .  The vacuum with unbroken $SO(32)$ has all open strings
terminating at the same fixed point.

\bigskip\noindent
{\it Heterotic - Type I Duality}
\par\nobreak\smallskip\nobreak

We are now ready to try
 to relate the eleven-dimensional picture to the conjectured heterotic
- Type I duality of ten-dimensional theories with gauge group $SO(32)$.

What suggests a
connection is the following.  Consider the Type I superstring
on $\R^9\times \S^1$.  Its  $T$-dual is related, as we have discussed,
to the Type IIA theory on an $\R^9\times \S^1/\Z_2$ orbifold.
We can hope to   identify the Type IIA
theory on $\R^9\times \S^1/\Z_2$ with the $M$-theory on $\R^9\times \S^1/\Z_2
\times \S^1$, since in general one hopes to associate Type IIA theory
on any space $X$ with $M$-theory on $X\times \S^1$.

On the other hand,
we have interpreted the $E_8\times E_8$ heterotic string as $M$-theory
on $\R^{10}\times \S^1/\Z_2$, so the $M$-theory on $\R^9\times \S^1\times
\S^1/\Z_2$ should be the $E_8\times E_8$ theory on $\R^9\times \S^1$.

So we now have two ways to look at the $M$-theory on $X=\R^9\times \S^1/\Z_2
\times\S^1$.  (1) It is the Type IIA theory on $\R^9\times \S^1/\Z_2$
which is also the $T$-dual of the Type I theory on $\R^9\times \S^1$.
(2) After exchanging the last two factors so as to write $X$
as $\R^9\times \S^1\times \S^1/\Z_2$, the same theory should be
the $E_8\times E_8$ heterotic string on $\R^9\times \S^1$.
So it looks like we can predict a relation between the Type I
and heterotic string theories!

This cannot be right in the form stated, since the model in (1) has
gauge group $SO(16)\times SO(16)$, while that in (2) has gauge group
$E_8\times E_8$.
Without really understanding the $M$-theory, we cannot
properly explain what to do, but pragmatically the most sensible course
is to turn on a Wilson line in theory (2), breaking $E_8\times E_8$
to $SO(16)\times SO(16)$.

At this point, it is possible that (1) and (2) are equivalent
(under exchanging the last two factors in $\R^9\times \S^1/\Z_2\times \S^1$).
The equivalence does not appear, at first sight, to be a known
equivalence between string theories.  We can relate it to a known
equivalence by making a $T$-duality transformation on each side.
In (1), a $T$-duality     transformation will convert to the
Type I theory on $\R^9\times \S^1$ (in its $SO(16)\times SO(16)$ vacuum).
In (2), a $T$-duality transformation will convert to
an $SO(32)$ heterotic string with $SO(32)$ spontaneously broken
to $SO(16)\times SO(16)$.\foot{$R\to 1/R$ symmetry, with $R$ the radius
of the circle in $\R^9\times \S^1$, maps the heterotic string vacuum with
unbroken $SO(32)$ to itself, and maps the heterotic string vacuum
 with unbroken
$E_8\times E_8$ to itself, but maps a heterotic string vacuum with
$E_8\times E_8$ broken to $SO(16)\times SO(16) $ to a heterotic string vacuum
with $SO(32)$ broken to $SO(16)\times SO(16)$.  This follows from
facts such as those explained in \refs{\narain,\ginsparg}.}
At this point, theories (1) and (2) are Type I and heterotic $SO(32)$
theories (in  their respective $SO(16)\times SO(16) $ vacua), so we can
try to compare them using the conjectured heterotic - Type I duality.
It turns out that this acts in the expected way, exchanging the last
two factors in $X=\R^9\times \S^1/\Z_2\times \S^1$.

Since the logic may seem convoluted, let us recapitulate.
On side (1), we start with the $M$ theory on $\R^9\times \S^1/\Z_2\times
\S^1$, and interpret
is as the $T$-dual of the Type I theory on $\R^9\times \S^1$.
On side (2), we start with the
$M$-theory on $\R^9\times\S^1\times \S^1/\Z_2$, interpret it as
the $E_8\times E_8$ heterotic string on $\R^9\times \S^1$ and
(after turning on a Wilson line) make a $T$-duality transformation
to convert the gauge group to $SO(32)$.  Then we compare (1) and (2)
using heterotic - Type I duality, which gives the same relation that
was expected from the eleven-dimensional point of view.
One is still left wishing that one understood better the meaning
of $T$-duality in the $M$-theory.

We hope that this introduction will make the computation below
easier to follow.

\bigskip\noindent
{\it Side (1)}
\par\nobreak\smallskip\nobreak

We start with the $M$-theory on $\R^9\times \S^1/\Z_2\times \S^1$,
with radii $R_{10}$ and $R_{11}$ for the two circles.
We interpret this as the Type IA theory, which is the $\Z_2$ orbifold of the
Type II theory on $\R^9\times \S^1$, a $T$-dual of the Type I theory in a
vacuum with unbroken $SO(16)\times SO(16)$.

The relation between $R_{10}$ and $R_{11}$ and the Type IA
parameters (the ten-dimensional string coupling $\lambda_\IA$ and radius
$R_\IA$ of the $\S^1$) can be computed by comparing the low energy actions
for the supergravity multiplet.  The computation and result are
as in \witten:%
\foot{The second equation is equivalent to the Weyl rescaling
$g_{10,M}=\lambda_\IIA^{-2/3}g_{10,\IIA}$ obtained in \witten\ between the
ten-dimensional metrics as measured in the $M$-theory or Type IIA.  Also,
by $\lambda_\IA$ we refer to the {\it ten-dimensional\/} string coupling
constant; similar convention is also used for all other string theories
below.}
\eqn\hucu{\eqalign{R_{11} & = \lambda_\IA^{2/3} \cr
                   R_{10} & = \frac{R_\IA}{\lambda_\IA^{1/3}}.\cr}}
Now we make a $T$-duality transformation to an ordinary Type I
theory (with unbroken gauge group $SO(16)\times SO(16)$),
by the standard formulas $R_\I=1/R_\IA$,
$\lambda_\I=\lambda_\IA/R_\IA$.
So
\eqn\gugu{\eqalign{R_{11}& = \frac{\lambda_\I^{2/3}}{R_\I^{2/3}}\cr
                   R_{10}& = \frac{\lambda_\I^{1/3}}{R_\I^{2/3}}.\cr}}

\bigskip\noindent
{\it Side (2)}
\par\nobreak\smallskip\nobreak

Now we start with the $M$-theory on $\R^9\times \S^1\times\S^1/\Z^2$,
with the radii of the last two factors denoted as $R_{10}'$ and $R_{11}'$.
This is hopefully related to the $E_8\times E_8$ heterotic string
on $\R^9\times \S^1$, with the $M$-theory parameters being related
to the heterotic string coupling $\lambda_{E_8}$ and radius $R_{E_8}$ by
formulas
\eqn\uhucu{\eqalign{R'_{11} & = \lambda_{E_8}^{2/3} \cr
                    R'_{10} & = \frac{R_{E_8}}{\lambda_{E_8}^{1/3}}\cr}}
just like \hucu, and obtained in the same way.
After turning on a Wilson line and making a $T$-duality transformation
to an $SO(32)$ heterotic string, whose parameters $\lambda_h,R_h$
are related to those
of the $E_8\times E_8$ theory by the standard $T$-duality relations
$R_h=1/R_{E_8}$, $\lambda_h=\lambda_{E_8}/R_{E_8}$, we get the analog
of \gugu,
\eqn\ugugu{\eqalign{R_{11}' & = \frac{\lambda_h^{2/3}}{R_h^{2/3}} \cr
                    R_{10}' & = \frac{1}{\lambda_h^{1/3}R_h^{2/3}}.\cr}}

\bigskip\noindent
{\it Comparison}
\par\nobreak\smallskip\nobreak

Now we compare the two sides via the conjectured $SO(32)$ heterotic - Type I
duality according to which these  theories coincide with
\eqn\compon{\eqalign{\lambda_h & = {1\over \lambda_\I} \cr
                        R_h    & = {R_\I\over \lambda_\I^{1/2}}.\cr}}
A comparison of \gugu\ and \ugugu\ now reveals that the relation
of $R_{10},R_{11}$ to $R'_{10},R'_{11}$ is simply
\eqn\ompon{\eqalign{ R_{10} & = R_{11}' \cr
                     R_{11} & = R_{10}'. \cr}}
So -- as promised -- under this sequence of operations,
the natural symmetry
in eleven dimensions becomes standard
heterotic - Type I duality.

\newsec{Comparison to Type II Dualities}

We have seen a close analogy between the dualities that involve heterotic and
Type I string theories and relate them to the $M$-theory, and the
corresponding Type II dualities that relate the Type IIA theory to
eleven dimensions and the Type IIB
theory to itself.  The reason for this analogy is of course that while the
Type II dualities are all related to the compactification of the $M$-theory
on $\R^9\times\T^2$, the heterotic and Type I dualities are related to the
compactification of the $M$-theory on a $\Z_2$ orbifold of $\R^9\times\T^2$.
It is the purpose of this section to make this analogy more explicit.

\fig{A section of the moduli space of compactifications of the $M$-theory on
$\R^9\times\T^2$.  By virtue of the $\Z_2$ symmetry between the two compact
dimensions, only the shaded half of the diagram
is relevant.}{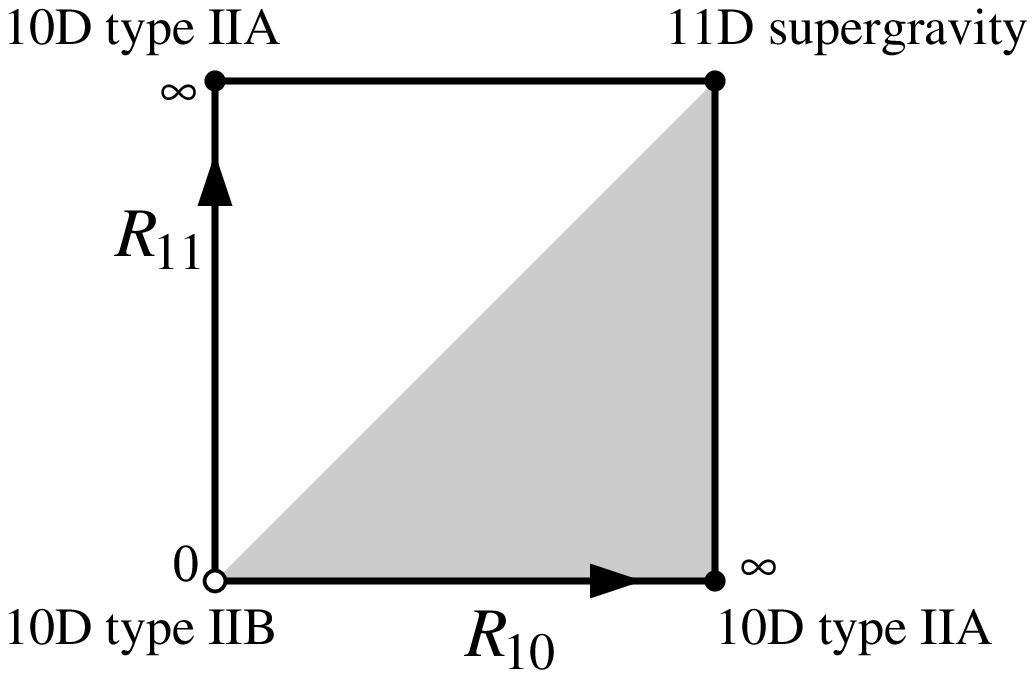}{3truein}

The moduli space of compactifications of the $M$-theory on a rectangular
torus is shown in Figure~1, following \refs{\aspinwall,\schwarz}.
Let us first recall how one can see the $SL(2,\Z)$ duality of the
ten-dimensional Type IIB theory in the moduli space of the $M$-theory on
$\R^9\times\T^2$.  The variables of the Type IIB theory on $\R^9\times\S^1$
are related to the compactification radii of the $M$-theory on
$\R^9\times\T^2$ by
\eqn\reloee{\eqalign{\lambda_\IIB&=\frac{R_{11}}{R_{10}},\cr
R_\IIB&=\frac{1}{R_{10}R_{11}^{1/2}}.\cr}}
The string coupling constant $\lambda_\IIB$ depends on the shape of the
two-torus of the $M$-theory, but not on its area.  As we send the radius
$R_\IIB$ to infinity to make the Type IIB theory ten-dimensional at fixed
$\lambda_\IIB$, the radii $R_{10}$ and $R_{11}$ go to zero.  So, the
ten-dimensional Type IIB theory at arbitrary string coupling corresponds to
the origin of the moduli space of the $M$-theory on $\R^9\times\T^2$ as shown
in Figure~1.  The $SL(2,\Z)$ duality group of the ten-dimensional Type IIB
theory can be identified with the modular group acting on the $\T^2$
\refs{\schwarz,\aspinwall}.

Similarly, the region of the moduli space where only one of the radii
$R_{10}$ and $R_{11}$ is small corresponds to the weakly coupled Type IIA
string theory.  When both radii become large simultaneously, the Type IIA
string theory becomes strongly coupled, and the low energy physics of the
theory is described by eleven-dimensional supergravity \witten .

Now we can repeat the discussion for the heterotic and Type I theories.  The
moduli space of the $M$-theory compactified on
$\R^9\times\S^1/\Z_2\times\S^1$ is sketched in Figure~2.
\fig{A section of the moduli space of compactifications of the $M$-theory
on $\R^9\times\S^1\times\S^1/\Z_2$.  Here $R_{10}$ is the radius of $\S^1/
\Z_2$.}{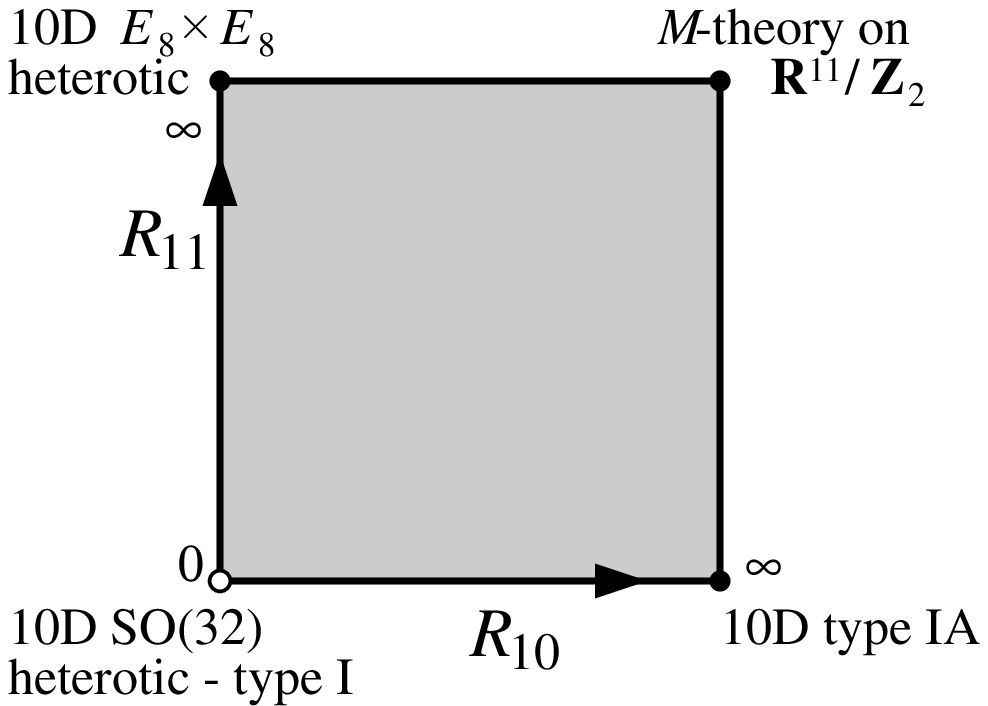}{3truein}
In the previous sections we discussed the relations between the
Type IA  theory on $\R^9\times\S^1/\Z_2$, the Type I theory
on $\R^9\times \S^1$, and the $M$-theory on
$\R^9\times\S^1/\Z_2\times\S^1$.  This relation leads to the following
expression for the Type I variables in terms of the variables of the
$M$-theory,
\eqn\inviee{\eqalign{\lambda_\I&=\frac{R_{11}}{R_{10}},\cr
                           R_\I&=\frac{1}{R_{10}R_{11}^{1/2}}.\cr}}
The string coupling constant $\lambda_\I$ depends only on the shape of the
two-torus of the $M$-theory.  By the same reasoning as in the Type IIB case,
the ten-dimensional Type I theory at arbitrary string coupling $\lambda_\I$
is represented by the origin of the moduli space, which should be -- in both
cases -- more rigorously treated as a blow-up.

We have related the $SO(32)$ heterotic string is related to variables of
the $M$-theory by
\eqn\invhee{\eqalign{\lambda_h&=\frac{R_{11}'}{R_{10}'},\cr
                           R_h&=\frac{1}{R_{10}'R_{11}'{}^{1/2}}.\cr}}
Using heterotic - Type I duality, which simply exchanges the two radii,
\eqn\ihduaee{\eqalign{R_{10}'&=R_{11},\cr
                     R_{11}'&=R_{10},\cr}}
we can express this relation in terms of $R_{10}$ and $R_{11}$,
\eqn\invhsee{\eqalign{\lambda_h&=\frac{R_{10}}{R_{11}},\cr
                            R_h&=\frac{1}{R_{11}R_{10}^{1/2}}.\cr}}
Just like the ten-dimensional Type I theory, the ten-dimensional $SO(32)$
heterotic theory at all couplings corresponds to the origin of the moduli
space.  The heterotic - Type I duality maps one of these theories at strong
coupling to the other theory at weak coupling, and vice-versa.

Now we would like to understand the regions of the moduli space where at
least one of the radii $R_{10}$, $R_{11}$ is large.  Recall that $R_{10}$ and
$R_{11}$  -- as measured in the $M$-theory -- are related to the string
coupling constants by
\eqn\sccee{\eqalign{R_{10}&=\lambda_{E_8}^{2/3},\cr
                    R_{11}&=\lambda_\IA^{2/3}.\cr}}
If one of the radii is large and the other one is small, the natural
description of the physics at low energies is in terms of the weakly-coupled
Type IA or $E_8\times E_8$ heterotic string theory.  Conversely, as we go to
the limit where both $R_{10}$ and $R_{11}$ are large, both string theories
are strongly coupled, and the low energy physics is effectively described by
the $M$-theory.

\bigskip\noindent
{\it Comparison of the Spectra}
\par\nobreak\smallskip\nobreak

We can gain some more insight into the picture by looking at some physical
states of the $M$-theory and interpreting them as states in different
weakly-coupled string theories.

A particularly natural set of states in the $M$-theory on $\R^9\times\T^2$ is
given by the Kaluza-Klein (KK) states of the supergravity multiplet, that is
the states carrying momentum in the tenth and eleventh dimension, along
with the wrapping modes of the membrane.  As measured in the $M$-theory,
these states have masses
\eqn\massmee{M^2=\frac{\ell^2}{R_{10}^2}+\frac{m^2}{R_{11}^2}+n^2R_{10}^2
R_{11}^2}
for certain values of $m,n,\ell$.

We are of course interested in states of the $M$-theory on
$\R^9\times\S^1/\Z_2\times\S^1$.  In order to get the states that survive on
the orbifold, we must project \massmee\ to the $\Z_2$ invariant sector.
Schematically, the orbifold group acts on the states with the quantum numbers
of \massmee\ as follows:
\eqn\orbactee{|\ell,m,n\rangle\to\pm|-\ell,m,n\rangle.}
The action of the orbifold group on the KK modes follows directly from
its action on the  space-time coordinates.  The action on the membrane
wrapping modes indicates that the orbifold changes simultaneously the
space-time orientation as well as the world-volume orientation of the
membrane.

While $m$ and $n$ are conserved quantum numbers even in the orbifold,
$\ell$ is not.  Nevertheless, we include it in the discussion since
$\ell$ is approximately conserved in some limits.

If our prediction about the relation of the $M$-theory on $\R^9\times\S^1/Z_2
\times\S^1$ to the heterotic and Type I string theories is correct, the
stable states of the $M$-theory must have an interpretation in each of these
string theories.  The string masses of these states as measured by the Type I
observer are
\eqn\massiee{M_\I^2=\ell^2R_\I^2+\frac{m^2R_\I^2}{\lambda_\I^2}
+\frac{n^2}{R_\I^2}.}
The membrane wrapping modes can be identified with the KK modes of the Type I
string, while the unstable states correspond to unstable winding modes of the
elementary Type I string.  The membrane KK modes along the eleventh dimension
are non-perturbative states in the Type I theory.  These states can be
identified with winding modes of non-perturbative strings with tension
$T_2\propto\lambda_\I^{-1}$.  We will see below that this is simply the
solitonic heterotic string of the Type I theory \refs{\dab,\hull,\pw}.

Similarly, we can try to interpret the states in the $T$-dual, Type IA
theory.  A Type IA observer will measure the following masses of the states:
\eqn\massiaee{M_\IA^2=\frac{\ell^2}{R_\IA^2}+\frac{m^2}{\lambda_\IA^2}+n^2
R_\IA^2.}
As required by $T$-duality, the membrane wrapping modes correspond to the
string winding modes. The unstable states correspond to the KK modes of the
Type IA closed string; they are unstable because the tenth component of the
momentum is not conserved in the Type IA theory.  The stable KK states of the
$M$-theory correspond to non-perturbative Type IA states.  These Type IA
states can be identified with the zero-branes (alias extremal black holes) of
the Type IIA theory in ten dimensions.  Notice that under the $\Z_2$ orbifold
action, the quantum number that corresponds to the extremal black hole states
is conserved, and the zero-brane states survive the orbifold projection.

In the $E_8\times E_8$ heterotic theory, the masses of our states are given by
\eqn\masshhee{M_{E_8}^2=\frac{m^2}{R_{E_8}^2}+n^2R_{E_8}^2.}
(We here omit $\ell$, as the unstable states it labels have no clear
interpretation for the weakly coupled heterotic string.)  The stable KK modes
along the eleventh dimension in the $M$-theory can be interpreted as the KK
modes along the tenth dimension in the heterotic theory, while the membrane
wrapping modes are the winding modes of the heterotic string.

In the $SO(32)$ heterotic string, the masses are
\eqn\masshee{M_h^2=m^2R_h^2+\frac{n^2}{R_h^2}.}
Again, these are the usual momentum and winding states of the heterotic
string.
The formulas also make it clear that  -- as expected from the
heterotic - Type I duality -- the $m=1$ non-perturbative Type I string state
corresponds to the elementary heterotic string.

We already pointed out an analogy between the heterotic - Type I duality and
the $SL(2,\Z)$ duality of the Type IIB theory; now we actually see remnants
of the $SL(2,\Z)$ multiplet of Type IIB string states in the $SO(32)$
heterotic and Type I theories.  This can be best demonstrated when we
consider the weakly-coupled Type IIB theory, and look at the behavior of its
spectrum under the $\Z_2$ orbifold group that leads to the Type I theory.
The perturbative Type IIB string of the $SL(2,\Z)$ multiplet is odd under the
$\Z_2$ orbifold action, and so does not give rise to a stable string.  But a
linear combination of strings winding in opposite directions survives the
projection and corresponds to the elementary Type I closed string, which is
unstable but long-lived for weak coupling.  The $SL(2,\Z)$ Type IIB string
multiplet also contains a non-perturbative state that is even under $\Z_2$,
and we have just identified it with  the elementary heterotic string.  Upon
orbifolding, the original $SL(2,\Z)$ multiplet of Type IIB strings thus gives
rise to both Type I and the heterotic string.

\bigskip\noindent
{\it Twisted Membrane States}
\par\nobreak\smallskip\nobreak

The states we have discussed so far are analogous to untwisted states of
string theory on orbifolds.  The membrane world-volume is without boundary,
but the membrane Hilbert space is projected onto $\Z_2$-invariant states;
the $\Z_2$ simultaneously reverses the sign of $x^{11}$ and the membrane
orientation.

We must also add the twisted membrane states, which are analogous to
open strings in parameter space orbifolds of Type II string theory reviewed
above.  Just like the open strings of the Type IA theory, the twisted
membrane states have world-volumes with two boundary components, restricted
to lie at one of the orbifold fixed points, $x^{10}=0 $ and $x^{10}=\pi$.
Such a state might simply be localized near one of the fixed points (in which
case the description as a membrane state might not really be valid), or it
might wrap around $\S^1/\Z_2\times \S^1$ a certain number of times (in which
case the membrane description does make sense at least if the radii are
large).  The former states might be called twisted KK states, and should
include the non-abelian gauge bosons discussed in section two.  The latter
states will be called twisted wrapping modes.  The twisted states carry no
momentum in the orbifold direction.  Both the momentum $\tilde m$ in the
$\S^1$ direction and the wrapping number $\tilde n$ are conserved.

States of these twisted sectors have masses -- as measured in the $M$-theory
-- given by
\eqn\masstmee{M^2=\frac{\tilde m^2}{R_{11}^2}+\tilde n^2R_{10}^2R_{11}^2.}
Just as in the untwisted sector, one has to project out the twisted states
that are not invariant under the orbifold group action.

Again, the $\Z_2$-invariant twisted states should have a natural
interpretation in the corresponding string theories.  In the Type I theory,
the twisted states of the $M$-theory have masses
\eqn\masstiee{M_\I^2=\frac{\tilde m^2R_\I^2}{\lambda_\I^2}+\frac{\tilde
n^2}{R_\I^2}.}
At generic points of our moduli space, the twisted states carry non-trivial
representations of $SO(16)\times SO(16)$.  The twisted wrapping modes of the
membrane correspond to the KK modes of the open Type I string.  The twisted
KK modes of the $M$-theory are non-perturbative string states of the Type I
theory, with masses $\propto R_\I/\lambda_\I$; they are charged under the
gauge group, and should be identified with the charged heterotic soliton
strings of the Type I theory.

In the Type IA theory, we obtain the following mass formula:
\eqn\masstiaee{M_\IA^2=\frac{\tilde m^2}{\lambda_\IA^2}+\tilde n^2R_\IA^2.}
While the twisted wrappping modes of the $M$-theory correspond to the
perturbative winding modes of the Type IA open string, the twisted KK modes
show up in the Type IA theory as additional non-perturbative black-hole
states, charged under $SO(16)\times SO(16)$.

In the $E_8\times E_8$ heterotic theory,
\eqn\massthhee{M_{E_8}^2=\frac{\tilde m^2}{R_{E_8}^2}+\tilde n^2R_{E_8}^2.}
Both sectors are perturbative heterotic string states in non-trivial
representations of $SO(16)\times SO(16)$.  In the $SO(32)$ heterotic theory,
the corresponding masses are
\eqn\massthee{M_h^2=\tilde m^2R_h^2+\frac{\tilde n^2}{R_h^2},}
which is of course in accord with $T$-duality.

One can go on and analyze spectra of other $p$-branes.  Let us only notice
here that the space-time orbifold singularities of the $M$-theory on
$\R^9\times\S^1/\Z_2\times\S^1$ are intriguing $M$-theoretical analogs of
Dirichlet-branes of string theory.

\bigskip\bigskip\noindent
We would like to thank  C.V. Johnson, J. Polchinski, J.H. Schwarz, and
N. Seiberg for discussions.

\listrefs
\end